\documentclass[runningheads]{llncs}
\usepackage[T1]{fontenc}
\usepackage{graphicx}
\usepackage[utf8]{inputenc} 
\usepackage[T1]{fontenc}    
\usepackage{hyperref}       
\usepackage{url}            
\usepackage{booktabs}       
\usepackage{amsfonts}       
\usepackage{nicefrac}       
\usepackage{microtype}      
\usepackage{xcolor}         
\usepackage{amsmath} 
\usepackage{multirow}
\usepackage{cite}
\usepackage{bm}
\usepackage{ragged2e}
\usepackage{float}
\begin{document}
\title{Gyri vs. Sulci: Disentangling Brain Core-Periphery
Functional Networks via Twin-Transformer}
%
%

\author{Xiaowei Yu\inst{1} \and
Lu Zhang\inst{1} \and Haixing Dai\inst{2} \and Lin Zhao\inst{2}, Yanjun Lyu\inst{1}, Zihao Wu\inst{2}, Tianming Liu\inst{2}, Dajiang Zhu\inst{1} }
\authorrunning{F. Author et al.}
%
\institute{The University of Texas at Arlington, Arlington TX 76019, USA \and
University of Georgia, Athens, 30609, USA
}

\maketitle              
\begin{abstract}
The human cerebral cortex is highly convoluted into convex gyri and concave sulci. It has been demonstrated that gyri and sulci are
significantly different in their anatomy, connectivity, and function: besides exhibiting opposite shape patterns, long-distance axonal fibers connected to gyri are much denser than those connected to sulci, and neural signals on gyri are more complex in low-frequency while sulci are more complex in high-frequency. Although accumulating evidence shows significant differences between gyri and sulci, their primary roles in brain function have not been elucidated yet. To solve this fundamental problem, we design a novel Twin-Transformer framework to unveil the unique functional roles of gyri and sulci as well as their relationship in the whole brain function. Our Twin-Transformer framework adopts two structure-identical (twin) Transformers to disentangle spatial-temporal patterns of gyri and sulci: one focuses on the information of gyri and the other is on sulci. The Gyro-Sulcal interactions, along with the tremendous but widely existing variability across subjects, are characterized in the loss design. We validated our Twin-Transformer on the HCP task-fMRI dataset, for the first time, to elucidate the different roles of gyri and sulci in brain function. Our results suggest that gyri and sulci could work together in a core-periphery network manner, that is, gyri could serve as core networks for information gathering and distributing, while sulci could serve as periphery networks for specific local information processing. These findings have shed new light on our fundamental understanding of the brain’s basic structural and functional mechanisms.

\keywords{Gyri and Sulci \and Core-Periphery \and Functional brain networks.}
\end{abstract}
\begin{figure}[ht]
\flushleft
\includegraphics[scale=0.23]{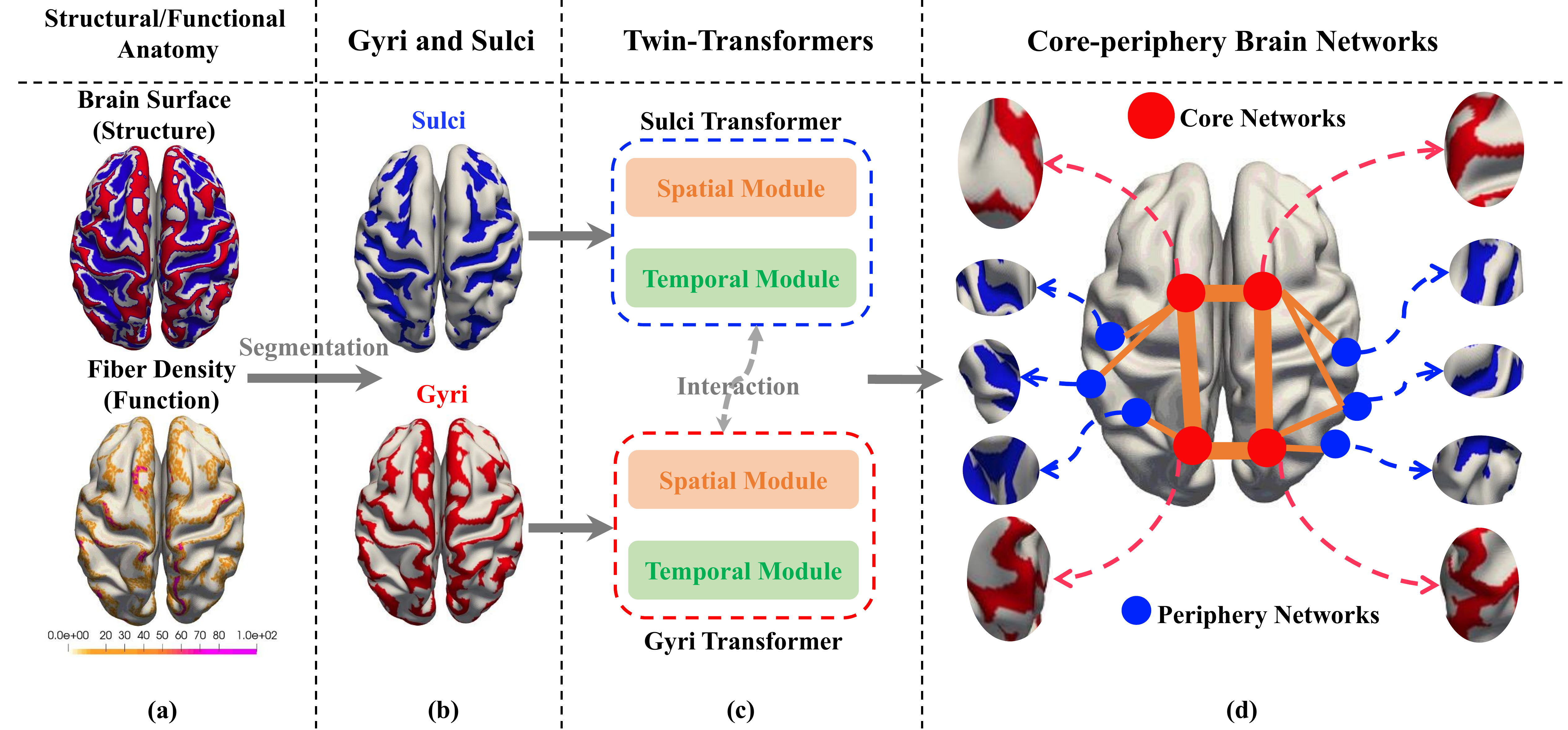}
\caption{Core-periphery brain networks in gyri and sulci. Part (a) is the brain structural and functional anatomy of the gyri and sulci. Part (b) is the segmentation of gyri and sulci. Part (c) is the proposed Twin-Transformer, where one is for gyri and the other is for sulci. Part (d) is the core-periphery brain network derived from the gyri and sulci, where gyri is the core network, and sulci is the periphery network.}
\label{fig_gyri_sulci}
\end{figure}

\section{Introduction}
Gyri and sulci are the standard morphological and anatomical nomenclature of the cerebral cortex (Figure \ref{fig_gyri_sulci}-a, b) and are usually defined in anatomical domains \cite{jiang2021fundamental}. Gyri and sulci serve as the basic building blocks to make up complex cortical folding patterns and are fundamental to realizing the brain’s basic structural and functional mechanisms. Numerous efforts have been devoted to understanding the function-anatomy patterns of gyri and sulci from various perspectives, including genetics\cite{richiardi2015correlated}, cell biology\cite{gertz2015neuronal}, and neuroimaging\cite{liu2019cerebral}. It has been demonstrated consistently that gyri and sulci are significantly different in their anatomy, connectivity, and function. Several studies\cite{fischl2000measuring, hilgetag2005developmental} found that the formation of gyri/sulci may be closely related to the micro-structure of white matter. For example, diffusion tensor imaging (DTI) derived long-distance axonal fibers connected to gyri are significantly denser than those connected to sulci (bottom of Figure \ref{fig_gyri_sulci}-a). That is, the long-distance fiber terminations dominantly concentrate on gyri rather than sulci, and interestingly, this phenomenon is evolutionarily preserved across different primate species. Meanwhile, using functional magnetic resonance imaging (fMRI), a few functional measurements that can directly reflect brain functional activities on gyri and sulci have been explored, such as functional BOLD signals \cite{liu2019cerebral}, correlation-based connectivity/interaction \cite{deng2014functional}, and spatial distribution of functional networks \cite{lv2014holistic}. Despite accumulating functional differences found between gyri and sulci, their basic roles as well as their relationship and interaction in the whole brain function have not been explored or elucidated yet.

To answer this fundamental question in brain science, we proposed a novel Twin-Transformer framework (Figure \ref{fig_gyri_sulci}-c) to explore and unveil the unique functional roles of gyri and sulci. Unlike traditional factorization-based approaches that assume linearity and independence, the Transformer attention mechanism is an ideal backbone to characterize, represent and reveal the complex and deeply buried patterns in the observed brain functional data. The whole framework is illustrated in Figure \ref{fig_TT}. Our Twin-Transformer framework adopts two structure-identical (twin) Transformers to model and disentangle spatial-temporal patterns of gyri and sulci: one focuses on the information of gyri and the other focuses on sulci. To model the complex 4D (spatial-temporal) fMRI data, within each transformer, we designed a spatial module and a temporal module to disentangle and extract the patterns in both spatial and temporal domains from the original fMRI signals. The two Transformers are connected and interact via a group of constraints on the commonality and specialty between the gyri and sulci. To effectively capture the Gyro-Sulcal interactions, as well as the tremendous and widely existing variability across individual brains, a novel Gyri-Sulci Commonality-Variability Disentangled Loss (GS-CV Loss) is proposed to guide the training process. After the model is well-trained, the spatial-temporal functional brain networks (FBNs) specific to gyri and sulci can be recovered. We validated our Twin-Transformer on one of the largest brain imaging datasets (HCP task-fMRI gray-ordinate dataset), for the first time, to elucidate the different roles of gyri and sulci in brain function. Our results suggest that gyri and sulci could work together in a core-periphery network manner (Figure \ref{fig_gyri_sulci}-d), that is, gyri could serve as core networks for information gathering and distributing in a global manner, while sulci could serve as periphery networks for specific local information processing. These findings have shed new light on our fundamental understanding of the brain's basic structural and functional mechanisms. The contributions of this paper are summarized as follows: 
\begin{itemize}
    \item We proposed a novel Twin-Transformer framework to disentangle the spatial-temporal patterns of the functional brain networks from fMRI datasets.
    \item We used the proposed method to represent and unveil the fundamental functional roles of the two basic cortical folding patterns: gyri and sulci.
    \item We found that gyri and sulci may work together in a Core-Periphery network manner: gyri serve as core networks for information gathering and distributing, while the sulci serve as periphery networks for specific local information processing.
\end{itemize}

\section{Methods}
\subsection{Gyri and Sulci Data Preparation}
\begin{figure}[t]
\centering
\includegraphics[scale=0.27]{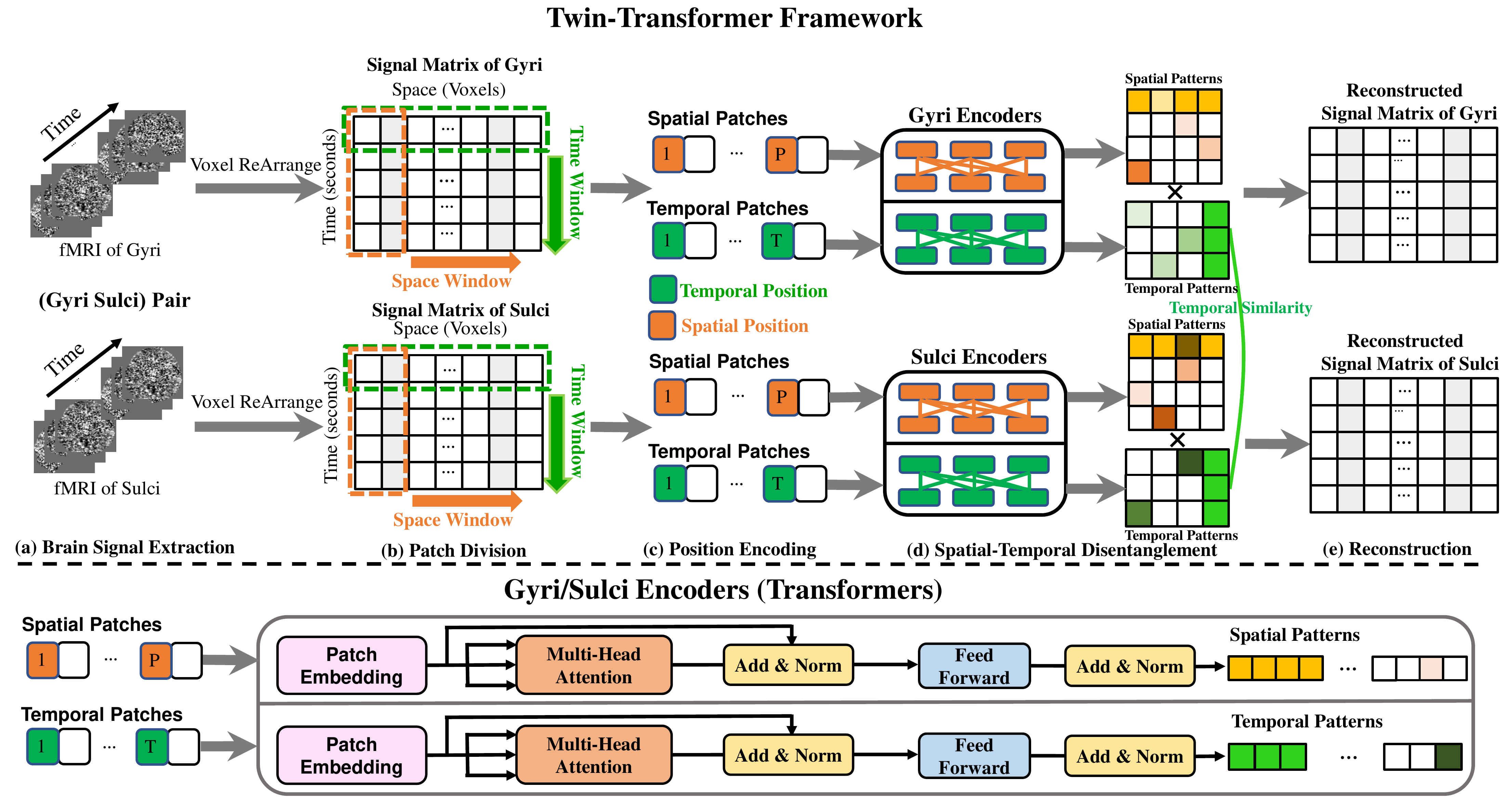}
\caption{Illustration of the proposed Twin-Transformer framework. Part (a) extracts the signals and rearranges them into a signal matrix according to ROIs. Part (b) shows the patch division of the gyri and sulci signal matrices. Part (c) is the position encoding for the spatial and temporal patches. Part (d) is the Twin-Transformer. Each transformer includes a spatial and temporal self-attention module for processing spatial and temporal patches, and the gyri transformer interacts with the sulci transformer under the loss design. Part (e) is the reconstruction of the gyri and sulci signal matrices from disentangled spatial and temporal patterns. The gyri/sulci encoders under the black dot line show the details of the Twin-Transformer.}
\label{fig_TT}
\end{figure}

In our experiments, we use the motor and working memory (WM) task fMRI (tfMRI) data of 545 subjects from the Human Connectome Project (HCP) dataset\cite{liu2019cerebral}. The publicly available preprocessed tfMRI data went through the minimal preprocessing pipelines specially designed for the HCP dataset \cite{Essen2013wu}. The preprocessed tfMRI imaging data are 4D imaging data, which consists of a time series of 3D images of the brain. For motor task-fMRI, each voxel contains a series of brain signals of length 284 and 405 for working memory task-fMRI. We rearrange the signals in each voxel into a 2D matrix. In this way, a 4D tfMRI imaging can be represented by a 2D matrix, where each row stores brain signals at each time step, and each column stores brain signals in a specific voxel (Figure \ref{fig_TT}-a,b). We normalized the brain signals to zero mean and unit variance. Since each subject of the preprocessed data has 59,412 voxels in standard grayordinate space, the column dimension of the 2D matrix is 59,412. To facilitate patch partition, we expand the space dimension according to our needs by adding zero vectors along the spatial dimension. For example, to disentangle the signal matrices into 50 spatial-temporal brain networks, the space dimension is extended from 59,412 to 59450 to divide the space dimension into 50 patches.  

\subsection{Twin-Transformer}
To reveal the common and variable patterns contained in the gyri and sulci, a novel Twin-Transformer framework is proposed, including a gyri transformer and a sulci transformer. The architecture of the Twin-Transformer is illustrated in Figure \ref{fig_TT}. There is a spatial and a temporal self-attention module in the gyri and sulci transformer for disentangling spatial and temporal patterns of gyri and sulci as shown in Figure \ref{fig_TT}-d. The structure of the sulci transformer is the same as the gyri transformer. For each input signal matrix, spatial patches are generated by shifting the spatial window along the space dimension, as illustrated by the orange arrow in Figure \ref{fig_TT}-b, while temporal patches are generated by shifting the temporal window along the time dimension, as shown in the green arrow. The gyri transformer generates spatial and temporal patterns of brain networks on gyri, while the sulci transformer generates spatial and temporal patterns of those on sulci.
 
Specifically, within the gyri and sulci transformer, the spatial self-attention module is designed to learn the latent representations of spatial features, and it focuses on the space dimension and takes non-overlapping spatial patches as tokens to build attention across the spatial variant patches and generate spatial patterns. It divides the input signal matrix into P non-overlapping patches by shifting the sliding window (orange dotted box following orange arrow) from left to right along the space dimension. The size of the sliding window can be adjusted according to the size of the input data. Each spatial patch contains complete temporal information of the focal brain region. The P patches correspond to P patterns of brain networks as predefined. Patches are used as tokens, and each token is first fed into a linear projection layer to obtain the latent representation 
$
        z_{i}\in \Re ^{1\times D_1 }
$
and then the learnable spatial positional embedding, 
$
        E _{i}^{s}\in \Re ^{1\times D_1}
$
is added to the representations of each input token.
The spatial transformer encoder can be formulated as:
\begin{equation}
    \begin{aligned}
        Spa(Z)= MLP ( MSA( LN( z_{1}^{s} || z_{2}^{s} || z_{3}^{s} || ... || z_{P}^{s} ) ))
    \end{aligned}
\end{equation}
where MSA() is the multi-head self-attention, MLP() represents multilayer perceptron, and LN() is layernorm.
$
        z_{i}^{s} = (z_{i} + E_{i}^{s})
$, 
$
i=1,2,...,P
$
and $||$ denotes the stack operation. $Spa(Z)\in P \times N$ is
the output of the spatial Transformer, where $P$ represents the number of brain networks, and $N$ is the number of voxels in the brain. $Spa(Z)$ models the activated status of voxels within
each brain network.

The temporal transformer is designed to learn the latent representations of temporal patterns of brain networks. The temporal self-attention module focuses on the temporal dimension and the non-overlapping temporal patches are used as tokens. Correspondingly, the temporal Transformer builds attention across the temporal variant patches and generates temporal features. Similar to the spatial transformer, by shifting the sliding window (green dotted box following green arrow) from top to bottom along the time dimension, $T$ non-overlapping temporal patches are generated. The size of the sliding window equals 1, hence the number of
patches equals the length of the brain signals. Each temporal patch contains information of all the voxels. After input embedding and positional embedding, each patch is represented by $z_{i}^{t} = (z_{i} + E_{i}^{t})$, $i=1,2,...,T$. The temporal self-attention module can be formulated as:
\begin{equation}
    \begin{aligned}
        Tem(Z)= MLP ( MSA( LN( z_{1}^{t} || z_{2}^{t} || z_{3}^{t} || ... || z_{P}^{t} ) ))
    \end{aligned}
\end{equation}
The outputs $Tem(Z)$ of the temporal self-attention module have a dimension of $Tem(Z)\in T \times P$, where $T$ represents the time length of the fMRI signals. $Tem(Z)$ represents the temporal patterns of the brain networks. Taking $Spa(Z)$ and $Tem(Z)$ together, we can obtain both the spatial and temporal patterns of each pair of gyri and sulci.

\subsection{
{Gyri and Sulci Commonality-Variability Disentangled Loss}
}
To simultaneously capture common and variable patterns in the gyri and sulci, a new gyri-sulci commonality-variability disentangled loss (GS-CV Loss) is proposed. There are three components in GS-CV Loss. The first one is the
signal matrix reconstruction loss. The whole framework is trained in a self-supervised manner to reconstruct the input signal matrix from the learned spatial and temporal patterns of gyri and sulci. This is crucial to ensure the learned spatial and temporal features can capture the complete spatial and temporal information of the input data. The reconstruction loss can be formulated as:

\begin{equation}
    \begin{aligned}
L_{reco}=\sum \left \| X - Spa(Z)\cdot Tem(Z) \right \|_{L1}
    \end{aligned}
\end{equation}
where $X$ is the input signal matrix, and we use the L1-norm to constrain the reconstruction of the input gyri and sulci pair. 

Despite the gyri and sulci having no overlap on the brain surface, different parts of the gyri and sulci may work synchronously and asynchronously in terms of temporal patterns. By constraining part of the temporal patterns of gyri and sulci to be similar and leaving others to learn freely, commonality and variability between gyri and sulci can be discovered. Therefore, the second component is the commonality constraint loss of temporal patterns between gyri and sulci, which aims to find the common temporal patterns between gyri and sulci. For this purpose, the learned temporal feature matrix is divided into the common part (the first $p$ columns) and the variable part (the remaining columns). The common and variable patterns can be learned by minimizing the difference between common parts of gyri and sulci and leaving the variable parts to learn freely. The commonality constraint on temporal features is formulated as:
\begin{equation}
    \begin{aligned}
L_{comm\_tem}=\sum Corr( \left \| Tem(Z_{1})[*,0:p] - Tem(Z_{2})[*,0:p] \right \| ) 
    \end{aligned}
\end{equation}
where $[*, 0:p]$ represents the first $p$ columns in $Tem(Z_{i})$, in which $\star$ means for each column, all the elements in the rows are included, and $Corr()$ is the Pearson correlation. In order to make spatial patterns distinct and limit the scale of temporal patterns from being arbitrarily large, we add a normalization on temporal patterns, which is formulated as:
\begin{equation}
    \begin{aligned}
L_{tem\_norm}= max(0, \frac{1}{P} ( \sum_{i=1}^{P}\left \| Tem(Z_i[*,i]) \right \|_{2} )-1 )
    \end{aligned}
\end{equation}

Combining the three parts, the GS-CV Loss can be formulated as:
\begin{equation}
    \begin{aligned}
GS{-}CV=\alpha L_{reco}+\beta L_{comm\_tem}+\gamma L_{tem\_norm}
    \end{aligned}
\end{equation}
where the regularization parameters $\alpha$, $\beta$, and $\gamma$ control the balance of different factors on the overall loss function.

\begin{figure}[ht]
\centering
\includegraphics[scale=0.085]{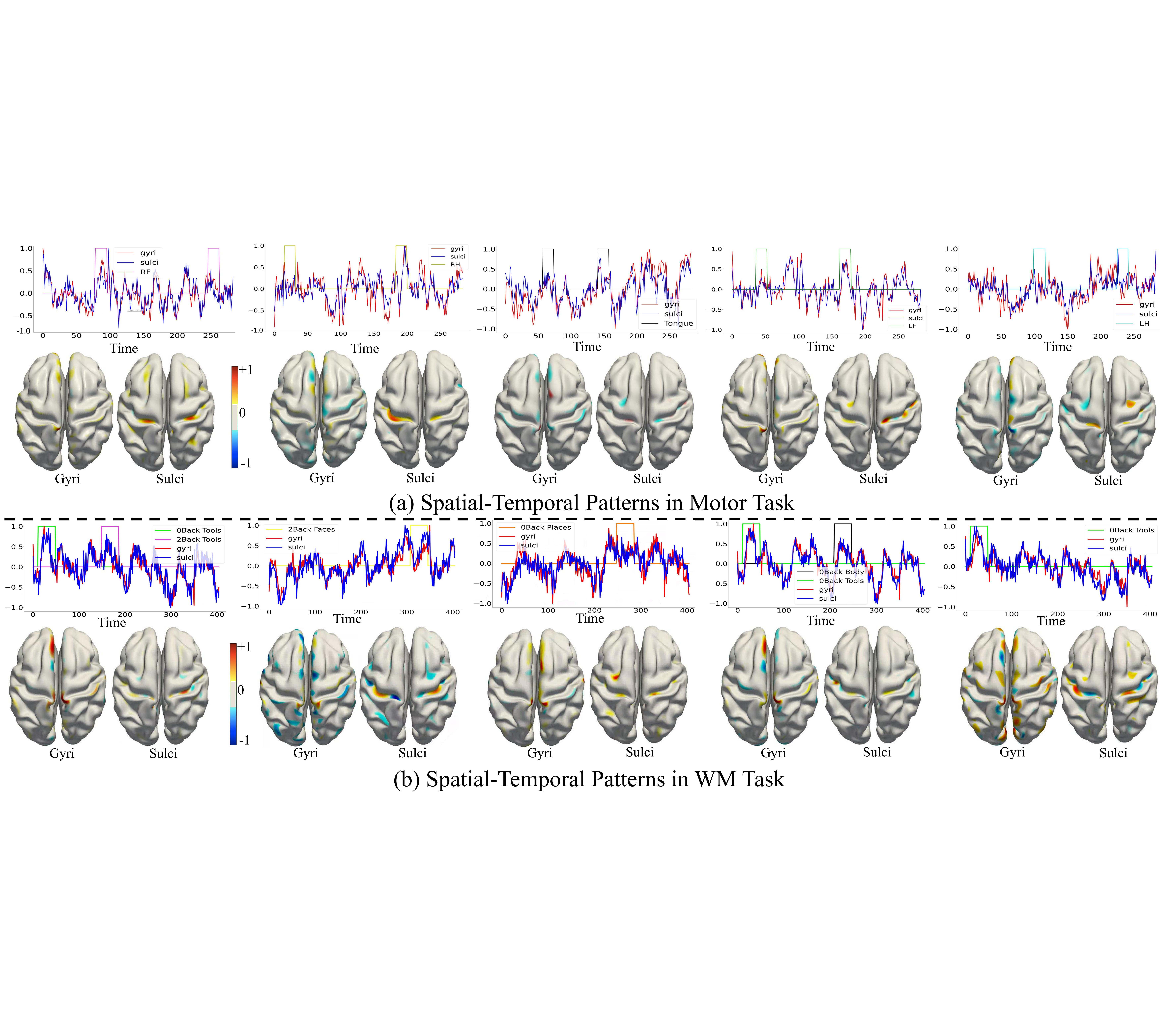}
\caption{Spatial-temporal functional brain networks disentangled from the motor and working memory tasks. These brain networks are from a randomly selected subject, separated by the black dot lines. The temporal patterns highly correlated to the specific task designs are presented with the corresponding stimulus.}
\label{tempspaBNs}
\end{figure}

\section{Results}
We applied our method to the HCP tfMRI of motor and working memory tasks. Using the fMRI signals from the gyri and sulci of each subject, as a paired input for Twin-Transformer, the output of the well-train model includes spatial-temporal patterns of gyri and sulci, where spatial patterns can be interpreted as FBNs, and the corresponding temporal patterns can be treated as the representative signals of each FBN. We implemented extensive experiments in different experimental settings, i.e., we set different numbers of spatial-temporal patterns, including [50, 100, 200]. We present the disentangled spatial-temporal patterns of gyri/sulci, core-periphery organized gyri/sulci, and ablation study of our proposed method in the following subsections. 

\subsection{{Spatial-Temporal Patterns of Gyri and Sulci}}
We first illustrate the disentangled spatial-temporal patterns of gyri/sulci. The results of both motor and working memory are from the experimental settings of 100 patterns, which consist of 50 common and 50 gyri/sulci-specific patterns. The spatial patterns are mapped to the brain surface, and after filtering by a preset threshold, we can obtain activated brain voxels (brain networks). We show five representative spatial-temporal functional brain networks of gyri/sulci from a randomly selected subject of the motor and working memory tasks in Figure \ref{tempspaBNs}. We can observe that multiple brain functional networks exist in the motor and WM tasks. For the motor task, there are five subtask designs, which are right hand(RH), left hand(LH), tongue, right foot(RF), and left foot(LF)\cite{Barch2013Function}. Five temporal patterns highly correlated with the specific subtask designs are shown in Figure \ref{tempspaBNs}-a. For the WM task, there are four explanatory variables(EVs), i.e., body, faces, places, and tools \cite{Barch2013Function}. Five disentangled spatial-temporal patterns are shown in Figure \ref{tempspaBNs}-b, in which temporal patterns highly correlated to EVs are presented with the stimuli. The disentangled spatial-temporal patterns of motor

\begin{figure}[H]
\centering
\includegraphics[scale=0.15]{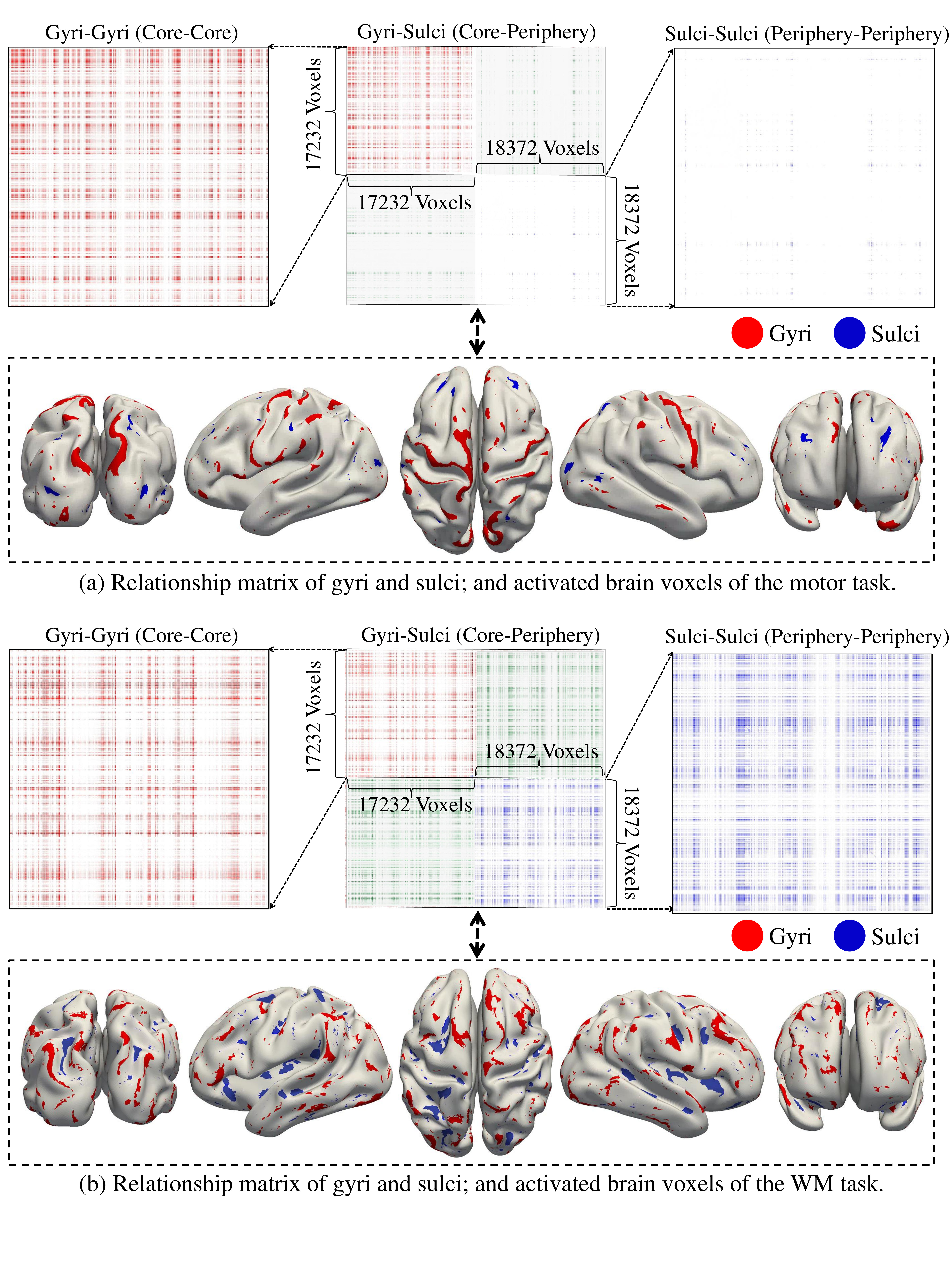}
\caption{ Core-periphery relationship between gyri and sulci in group-level analysis. Part (a): The relationship matrix and corresponding activated voxels of the motor task. Part (b): The relationship matrix and corresponding activated voxels of the WM task. For better visualization, we enlarge the gyri and sulci parts of the relationship matrix and display them on the left and right beside the entire matrix. The activated brain voxels are shown in five different perspectives. }
\label{cp}
\end{figure}

\noindent and WM tasks demonstrate that our proposed Twin-Transformer framework can effectively obtain the spatial-temporal brain networks of different subjects from fMRI data.

\subsection{{Core-Periphery Network in Gyri and Sulci}}

We can identify the activated brain voxels whose weights are consistently above a pre-defined threshold across all gyri- or sulci- derived spatial patterns. By connecting all the activated brain voxels, we construct a relationship matrix of gyri and sulci. Combining the spatial patterns of all the subjects and averaging the result, we can obtain the group-wise relationship matrix and the corresponding activated voxels on the brain surface, as shown in Figure \ref{cp}. There are 17,232 voxels for gyri and 18,327 voxels for sulci in this subject's gray-ordinate surface, so the dimension of the obtained relationship matrix is $35559 \times 35559 $, and $17232 \times 17232 $ and $ 18327 \times 18327 $ for the gyri part and sulci part, respectively. The results of both tasks, as shown in Figures \ref{cp}-a and \ref{cp}-b, include the relationship matrix (the upper part) and the activated voxels (the lower part). The middle one in the upper part shows the entire relationship matrix, and the sub-figures on the left and right highlight the connections within gyri and sulci voxels, which are located in the top-left and bottom-right of the entire relationship matrix. In general, the relation matrix is sparse, which means only a few regions (voxels) are involved in a specific task at the same time. The activated brain voxels from five different perspectives are shown in the lower part. As shown in Figure \ref{cp}, the activated brain voxels in the gyri-gyri section incline to form larger and connected blocks or clusters, while the activated brain voxels in the sulci-sulci section tend to assemble as smaller and scattered patterns. Besides, the brain networks of the motor task are concentrated on gyri, while working memory involves larger sulcal regions that host more high-level functions. It is worth noting that if the voxels are close in the relationship matrix, they also tend to be neighbors on the cortical surface. Therefore, after mapping the relationship matrix to the cortex, we see continuous regions on the brain surface forming gyri-based and sulci-based FBNs.

To examine and prove the concept of the Core-Periphery network of gyri and sulci, we compute the normalized independent probability (IP) $P_{GG}$, $P_{SS}$ and $P_{GS}$ for sub-matrices $A_{GG}$, $A_{SS}$, and $A_{GS}$ of the entire relationship matrix, which represents the interactions within gyri vertices (Core Network), sulci vertices (Periphery Network), and between gyri and sulci vertices (between Core and Periphery Networks). Independent probability\cite{cucuringu2016detection} is defined as the probability that there is an edge between any pairs of nodes in a given matrix, and it is an important measurement to indicate if the matrix or graph is organized as Core-Periphery pattern\cite{rombach2014core}. The independent probability and normalized independent probability are formulated as:

\begin{table}[ht]
\caption{The normalized independent probability of gyri and sulci brain networks under 100 patterns set, in which 50 common and 50 gyri/sulci specific patterns.}
\centering
\begin{tabular}{ccccccc}
\hline
\multirow{2}{*}{IP} & \multicolumn{3}{c}{ MOTOR} & \multicolumn{3}{c}{WM} \\
\cline{2-7} 
                   & 0.60  & 0.50  & 0.40  & 0.60  & 0.50   & 0.40   \\
\hline
$P_{GG}$ \quad &$0.70\pm{0.05}$ \quad   &  $0.66\pm{0.01}$ \quad   &  $0.46\pm{0.01}$ \quad  &$0.42\pm{0.13}$ \quad & $0.40\pm{0.11}$ \quad   &  $0.37\pm{0.08}$          \\
$P_{GS}$ \quad  &$0.14\pm{0.03}$  \quad  & $0.18\pm{0.01}$
\quad & $0.32\pm{0.02}$  \quad &  $0.31\pm{0.03}$   \quad  & $0.32\pm{0.02}$  \quad  &    $0.33\pm{0.02}$	        \\
$P_{SS}$ \quad  &$0.10\pm{0.02}$ \quad   & $0.15\pm{0.01}$  \quad   &  $0.22\pm{0.01}$  \quad  &  $0.27\pm{0.11}$  \quad  &$0.28\pm{0.09}$  \quad  & $0.30\pm{0.07}$           \\
\hline
\label{cp_value}
\end{tabular}
\end{table}

\begin{equation}\label{normalizedIP}
\begin{aligned}
I_{GG} = \frac{  1_{A_{GG}} }{ \lVert{A_{GG}} \rVert_{1} },
I_{GS} = \frac{  1_{A_{GS}} }{ \lVert{A_{GS}} \rVert_{1} },
I_{SS} = \frac{  1_{A_{SS}} }{ \lVert{A_{SS}} \rVert_{1} },\\
P_{GG} = I_{GG} / (I_{GG}+ I_{GS} + I_{SS}),\\
P_{GS} = I_{GS} / (I_{GG}+ I_{GS} + I_{SS}),\\
P_{SS} = I_{SS} / (I_{GG}+ I_{GS} + I_{SS}).
\end{aligned}
\end{equation}

\noindent where $1_{A_{GG}}$ represents the number of $1$s in the sub-matrix of gyri-gyri and ${\lVert \bullet \rVert}_{1}$ is the number of elements in this sub-matrix. The same procedure was applied to sub-matrices of core-periphery and
periphery-periphery, respectively. We set three different thresholds for filtering activated voxels to calculate the normalized independent probability, and the averaged results of 545 subjects are shown in Table \ref{cp_value}. The results show that $P_{GG}$ > $P_{GS}$ > $P_{SS}$, which confirms that our derived gyri/sulci brain functional networks have the core–periphery structure. Considering the results in Figure \ref{cp} and Table \ref{cp_value}, the core-periphery structure is more prominent in brain networks of the motor task than the counterparts appearing in the WM task.

\begin{table}[bp]
\caption{The normalized independent probability of gyri and sulci brain networks under 50 patterns set, in which 25 common and 25 gyri/sulci specific patterns.}
\centering
\begin{tabular}{ccccccc}
\hline
\multirow{2}{*}{IP} & \multicolumn{3}{c}{ MOTOR} & \multicolumn{3}{c}{WM} \\
\cline{2-7} 
                   & 0.60  & 0.50  & 0.40  & 0.60  & 0.50   & 0.40   \\
\hline
$P_{GG}$ \quad &$0.82\pm{0.07}$ \quad   & $0.62\pm{0.05}$  \quad   &  $0.44\pm{0.03}$ \quad  &$0.63\pm{0.06}$ \quad &$0.52\pm{0.02}$  \quad   &     $0.45\pm{0.06}$       \\
$P_{GS}$ \quad  &$0.11\pm{0.04}$  \quad  & $0.24\pm{0.02}$   \quad &$0.31\pm{0.01}$   \quad &  $0.28\pm{0.05}$   \quad  & $0.30\pm{0.02}$	 \quad  &    $0.32\pm{0.02}$         \\
$P_{SS}$ \quad  &$0.07\pm{0.04}$ \quad   &  $0.13\pm{0.04}$ \quad   & $0.24\pm{0.04}$   \quad  &  $0.10\pm{0.04}$  \quad  &$0.17\pm{0.05}$  \quad  &    $0.22\pm{0.03}$        \\
\hline
\label{cp_value_50}
\end{tabular}
\end{table}

\subsection{{Ablation Study}}
We conduct ablation studies to validate the effectiveness and robustness of the proposed Twin-Transformer framework and verify the core-periphery relationship of gyri and sulci under different experimental settings. We implement extensive experiments under different settings of the number of patterns, i.e., adjusting the sliding window size along the space dimension. The normalized independent probability of the relationship matrix under 50 and 200 patterns are shown in Tables \ref{cp_value_50} and \ref{cp_value_200}, respectively. We can observe that the normalized independent probability of the relationship matrix of gyri and sulci under different experimental settings and pre-defined thresholds all satisfy $P_{GG}$ > $P_{GS}$ > $P_{SS}$, which further demonstrates that the gyri-sulci functional brain networks are organized in a core-periphery manner.

\begin{table}[t]
\caption{The normalized independent probability of gyri and sulci brain networks under 200 patterns set, in which 50 common and 150 gyri/sulci specific patterns.}
\centering
\begin{tabular}{ccccccc}
\hline
\multirow{2}{*}{IP} & \multicolumn{3}{c}{ MOTOR} & \multicolumn{3}{c}{WM} \\
\cline{2-7} 
                   & 0.60  & 0.50  & 0.40  & 0.60  & 0.50   & 0.40   \\
\hline
$P_{GG}$ \quad &$0.80\pm{0.06}$ \quad   & $0.70\pm{0.04}$  \quad   &  $0.36\pm{0.05}$ \quad  &$0.42\pm{0.18}$ \quad &$0.41\pm{0.15}$  \quad   &     $0.39\pm{0.12}$       \\
$P_{GS}$ \quad  &$0.15\pm{0.01}$  \quad  & $0.18\pm{0.01}$   \quad &$0.33\pm{0.01}$   \quad &  $0.29\pm{0.06}$   \quad  & $0.31\pm{0.05}$	 \quad  &    $0.32\pm{0.04}$         \\
$P_{SS}$ \quad  &$0.05\pm{0.02}$ \quad   &  $0.10\pm{0.04}$ \quad   & $0.29\pm{0.04}$   \quad  &  $0.27\pm{0.16}$  \quad  &$0.28\pm{0.13}$  \quad  &    $0.29\pm{0.11}$        \\
\hline
\label{cp_value_200}
\end{tabular}
\end{table}

\section{Discussion and Conclusion}
In the brain science field, gyri and sulci are known to possess different structural, connectional, and functional characteristics. However, it is the first time that our Twin-Transformer is accurate enough to quantitatively differentiate gyri and sulci into core-periphery networks, which might suggest that the cerebral cortex is segregated into two fundamentally different functional units of gyri and sulci. This result has profound impacts on many aspects of basic, cognitive, and clinical neuroscience. Core-periphery network phenomena have been reported in many real-world networked systems such as transportation \cite{roth2012long}, social networks \cite{boyd2006computing}, and biological neural networks \cite{guillon2019disrupted}, among others, and our work here revealed and characterized such core-periphery patterns in a fine-grained manner on cortical gyri and sulci. Given that the graph structures, e.g., the relational graph of CNNs, of artificial neural networks in highly optimized deep learning models are more similar to those in biological neural networks \cite{You2020Graph}, it is reasonable to postulate that the core-periphery network structure discovered in human brains in this work could be potentially infused into the design of next-generation artificial neural networks in deep learning as prior knowledge or meaningful constraint, thus leading to brain-inspired artificial intelligence. 

There are limitations. First, we mainly focus on discovering the relationship between gyri and sulci at this moment and ignored the intermediate regions on the gyral wall that is between gyri and sulci. Second, we lack investigations into the brain regions highly related to specific subtasks, such as tongue movement, viewing faces, and others. These limitations are opportunities for future improvements. 

In summary, we proposed a novel data-driven Twin-Transformer framework and applied it to HCP gray-ordinate tfMRI dataset to characterize the roles of cortical gyri and sulci on the brain functional networks. With this framework, we can disentangle the spatial and temporal patterns from the brain signals of gyri and sulci, providing us the possibility to analyze the difference between gyri and sulci. The most important finding in this study is that we identified the core-periphery relationship between gyri and sulci, as well as the corresponding core-periphery brain networks. Our results show that core-periphery networks are broadly existing between gyri and sulci across different subjects and different tasks. Overall, our proposed Twin-Transformer contributes to a better understanding of the roles of gyri and sulci as core and periphery in brain architecture.

%
%
%
%

\end{document}